\documentclass[12pt]{article}
\usepackage{graphicx}
\usepackage{lineno}

\newcommand\pubnumber{SNSN-323-63}
\newcommand\pubdate{\today}

\def\genova{Istituto Nazionale di Fisica Nucleare\\
 Genova, ITALY}

\def\Title#1{\begin{center} {\Large #1 } \end{center}}
\def\Author#1{\begin{center}{ \sc #1} \end{center}}
\def\Address#1{\begin{center}{ \it #1} \end{center}}

\newcommand\pubblock{\rightline{\begin{tabular}{l} \pubnumber\\
         \pubdate  \end{tabular}}}
\newenvironment{Abstract}{\begin{quotation}  }{\end{quotation}}
\newenvironment{Presented}{\begin{quotation} \begin{center} 
             PRESENTED AT\end{center}\bigskip 
      \begin{center}\begin{large}}{\end{large}\end{center} \end{quotation}}

\def\beq{\begin{equation}}
\def\eeq#1{\label{#1}\end{equation}}
\def\eeqn{\end{equation}}

\def\beqa{\begin{eqnarray}}
\def\eeqa#1{\label{#1}\end{eqnarray}}
\def\eeqan{\end{eqnarray}}

\let\bar=\overbar

\def\Dslash{\not{\hbox{\kern-4pt $D$}}}
\def\dslash{\not{\hbox{\kern-2pt $\del$}}}

\def\msb{{\bar{\ssstyle M \kern -1pt S}}}

\begin{document}
\begin{titlepage}
\pubblock

\vfill
\Title{Latest ATLAS results from Run 2}
\vfill
\Author{Claudia Gemme}
\Address{\genova}
\Author{on behalf of the ATLAS Collaboration}
\vfill
\begin{Abstract}
After the first LHC  long shutdown with upgrades to the machine and the detectors, since 2015 the ATLAS experiment 
 recorded more than 30~fb$^{-1}$ of integrated
luminosity of $pp$ collision data at 13~TeV centre-of-mass energy.  
The data collected to date, the detector and physics performance, and measurements of  Standard Model
processes are reviewed briefly before summarising the latest ATLAS results in the Brout-
Englert-Higgs sector, where substantial progress has been made since the discovery. Searches for physics phenomena beyond the Standard Model are also summarized. 
These proceedings reflect only a brief summary of the material presented at the conference.
\end{Abstract}
\vfill
\begin{Presented}
9th International Worshop on top quark physics\\
Olomouc, Czech Republic,  September 19--23, 2016
\end{Presented}
\vfill
\end{titlepage}
\def\thefootnote{\fnsymbol{footnote}}
\setcounter{footnote}{0}
%

\section{Introduction}

After the first succesful data taking period in 2010-2012 (Run 1), the ATLAS detector~\cite{ATLASJinst} has been significantly improved during the LHC long
shutdown in 2013-2014 with a new beam pipe and a 4th silicon pixel layer (IBL) at 3.3 cm from the interaction point; improvements in the magnetic and cryogenic systems; consolidation and repairs of all subsdetectors; upgrade in the trigger and data acquisistion systems  increasing the  maximum first level hardware rate from 75~kHz to 100~kHz and
merging the two software trigger levels.
 The ATLAS experiment has had a successful start of $pp$ data taking at 13~TeV (Run 2) with 
3.9~fb$^{-1}$  collision data  recorded in 2015, with a DAQ efficiency of 92\%. In 2016, 27.1~fb$^{-1}$ of $pp$ collision data were recorded by the time of this conference, with very high efficiency. 
The peak luminosity delivered by LHC was 1.37 $\times$ 10$^{34}$ cm$^{-2}$ s$^{-1}$, greater than the design value of 1 $\times$ 10$^{34}$ cm$^{-2}$ s$^{-1}$.
The status of the detector is  excellent, with close to 100\% of readout
channels available across all sub-detectors.
\begin{figure}[htb]
\centering
\includegraphics[width=0.49\textwidth]{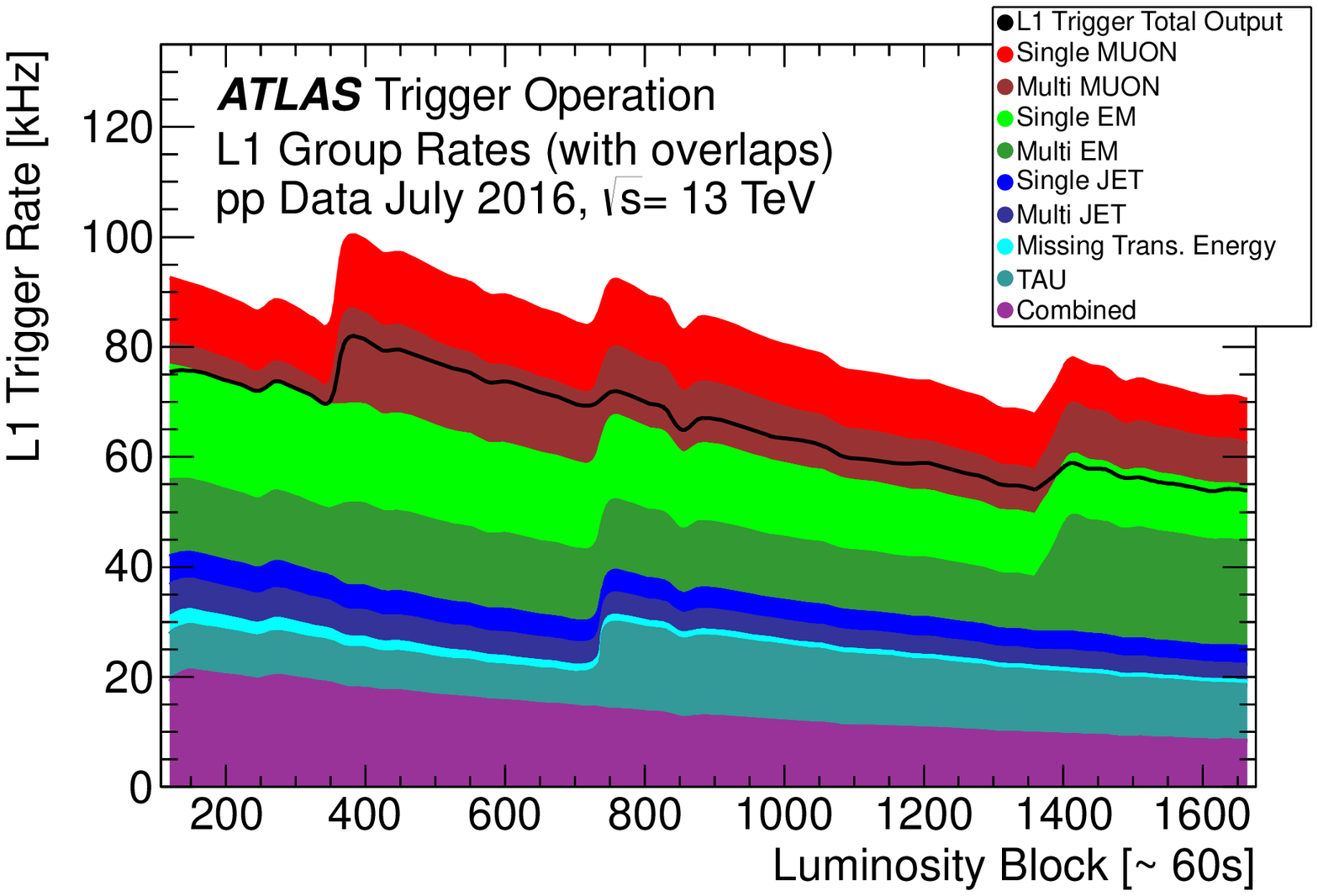}
\includegraphics[width=0.49\textwidth]{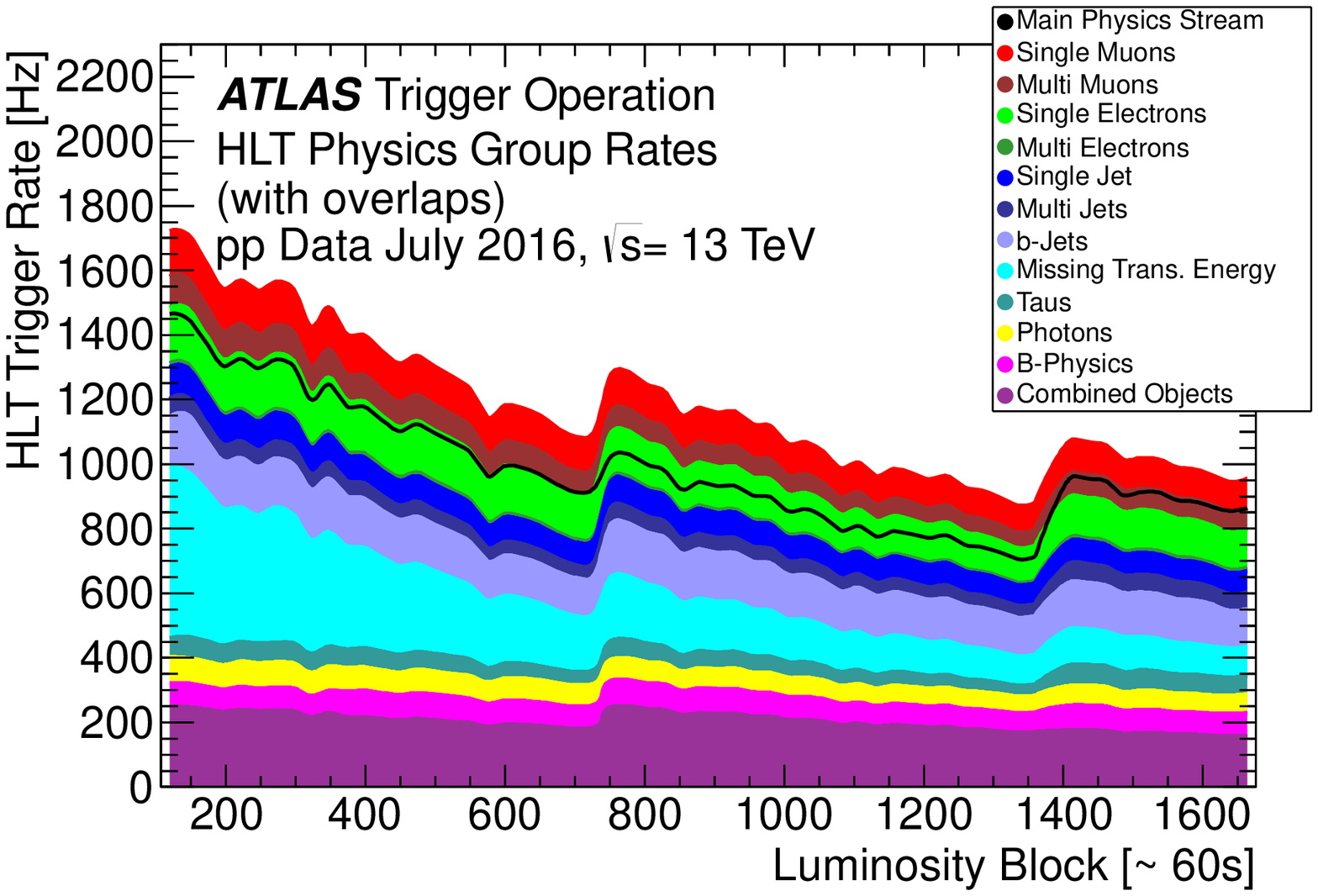}
\caption{
Left: Physics trigger group rates at the first trigger level (L1) for one fill as a function of the luminosity block
number. Each luminosity block corresponds on average to 60~s. The fill was taken in July 2016 with a
peak luminosity of $1.02\times$ 10$^{34}$ cm$^{-2}$ s$^{-1}$  and a peak pile-up of  35.
Presented are the rates of the individual L1 trigger groups for various L1 trigger physics objects. Overlaps
are accounted for in the total output rate, but not in the individual groups, leading to a higher recording rate
compared to the total L1 output rate. Right: Same as on the left, but physics trigger group rates at the High Level Trigger (HLT). \cite{TriggerPublicTwiki} 
}
\label{fig:L1_Trigger}
\end{figure}

The improvements in the event selection and readout systems have allowed  facing the  2016 conditions,
with a complex trigger menu designed to meet varied physics, monitoring and
performance requirements.  
 As can be seen in Fig.~\ref{fig:L1_Trigger} for a typical LHC fill, most of the bandwidth of the first level hardware (L1)  and  High Level Trigger software (HLT) rates is still given to generic triggers, such as single isolated leptons, complemented by multi-object triggers and triggers dedicated to specific analyses. The average physics output rate is  $\sim$1~kHz, keeping the single leptons threshold at  24-26 GeV and the missing energy  threshold at 90-110 GeV.

A detailed understanding of the detector performance is essential for the production of high
quality results. In particular, as the  mean number of interactions per crossing (pile-up) almost doubled in 2016 with respect to  2015, extensive work has been done to reduce its impact on the reconstruction performance  of basic objects such as leptons, $b$-jets, missing energy, etc.

\section{Results on Standard Model Physics}

\begin{figure}[htb]
\centering
\includegraphics[width=0.6\textwidth]{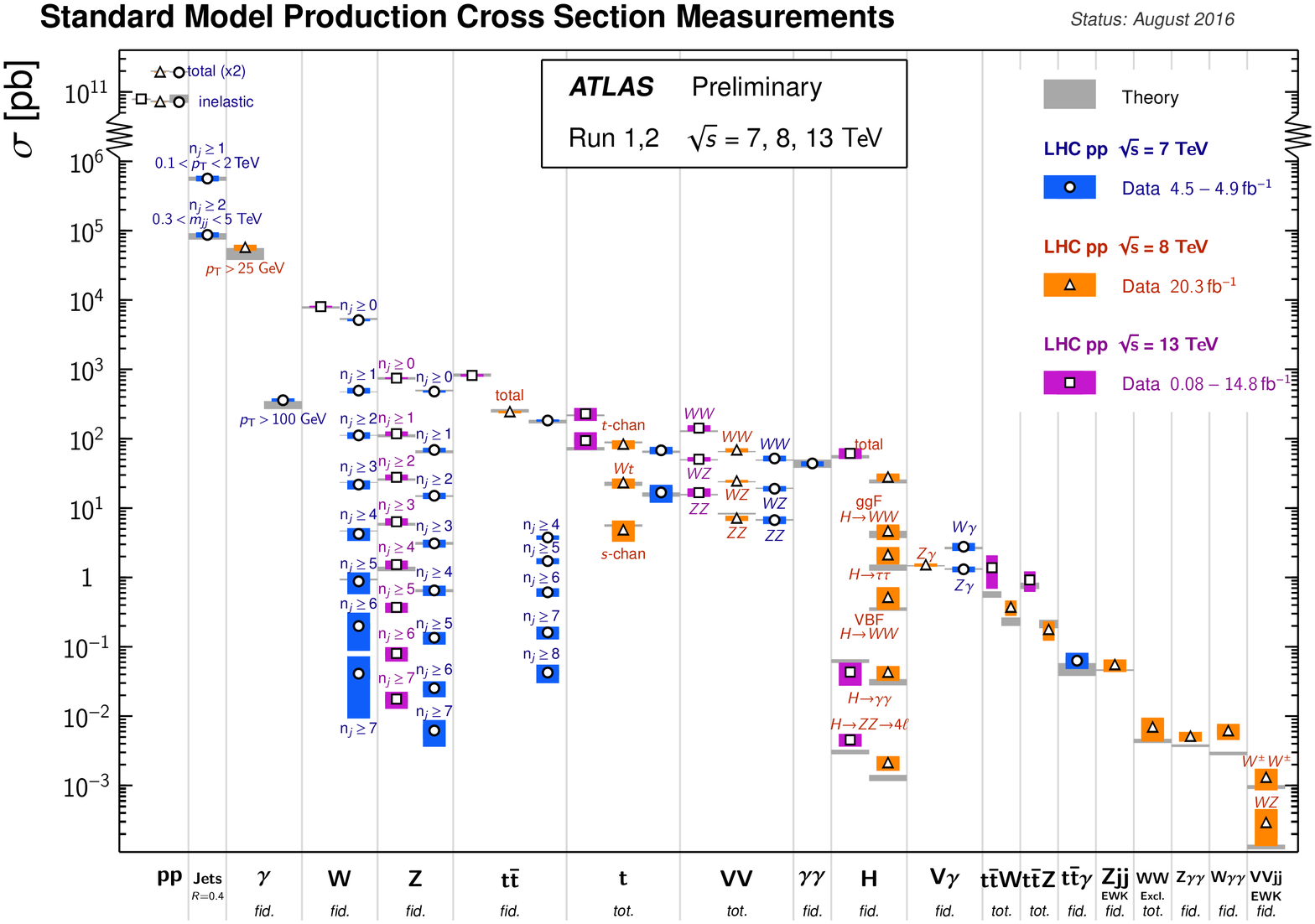}
\caption{Overview of cross-section measurements of selected Standard Model processes compared to the corresponding theoretical expectations~\cite{StandardModelPublicResults}. All theoretical expectations were calculated at NLO or higher.}
\label{fig:SM_Overview}
\end{figure}

A detailed understanding of the Standard Model (SM) processes is essential for the ATLAS physics program. 
While looking for possible deviations from SM predictions, they represent a key
ingredient for the description of the backgrounds and Monte Carlo models in the new
physics searches, which are pushing into increasingly intricate event signatures. 
An overview of such cross-section measurements  is shown in Fig.~\ref{fig:SM_Overview}. \\

The  measurements of $Z$ production in association with jets~\cite{ATLAS-CONF-2016-046} and massive diboson production~\cite{arXiv:1606:04017} were briefy mentioned at the conference,  and the interested reader is referred to the ATLAS results web page,
as well as to specific references.
These are just two examples demonstrating as accurate predictions from different Monte Carlo generators are needed  to  face the challenge of the precision of  LHC data.

\subsection{Higgs Boson measurements}
%
In July 2012, the ATLAS and CMS collaborations announced the discovery of a Higgs boson~\cite{ATLASHiggs1, CMSHiggs1} using
$pp$ collision data collected at centre-of-mass energies $\sqrt{s} = 7$~TeV and 8 TeV at the LHC. Using the full Run 1 statistics, 
ATLAS and CMS have summarized their measurement in combined legacy papers~\cite{ACmass, ACcouplings}. 
The Higgs boson mass is measured   in the $H\rightarrow\gamma\gamma$ and $H\rightarrow ZZ^{*} \rightarrow 4l$  decay channels. The results are obtained from a simultaneous fit to the reconstructed invariant mass peaks in the two channels and for the two experiments. The measured masses from the individual channels and the two experiments are found to be consistent among themselves. The combined measured mass of the Higgs boson is $m_H = 125.09 \pm 0.21\mathrm{(stat.)} \pm 0.11 \mathrm{(syst.)}$~GeV~\cite{ACmass}.
Combined ATLAS and CMS measurements of the Higgs boson production and decay rates,
as well as constraints on its couplings to vector bosons and fermions, are summarized in ~\cite{ACcouplings}.
The combined signal yield relative to the Standard Model prediction is measured
to be $1.09 \pm 0.11$. The combined measurements lead to observed significances for the
vector boson fusion production process and for the  $H\rightarrow\tau\tau$ decay of 5.4 and 5.5 standard deviations, respectively. 
Most couplings measurements are consistent with the SM predictions within 2$\sigma$. The largest
observed deviation is the ratio $\sigma_{ttH}/\sigma_{ggF}$ at 3.0 standard deviations to the SM.

\begin{figure}[htb]
\centering
\includegraphics[width=0.6\textwidth]{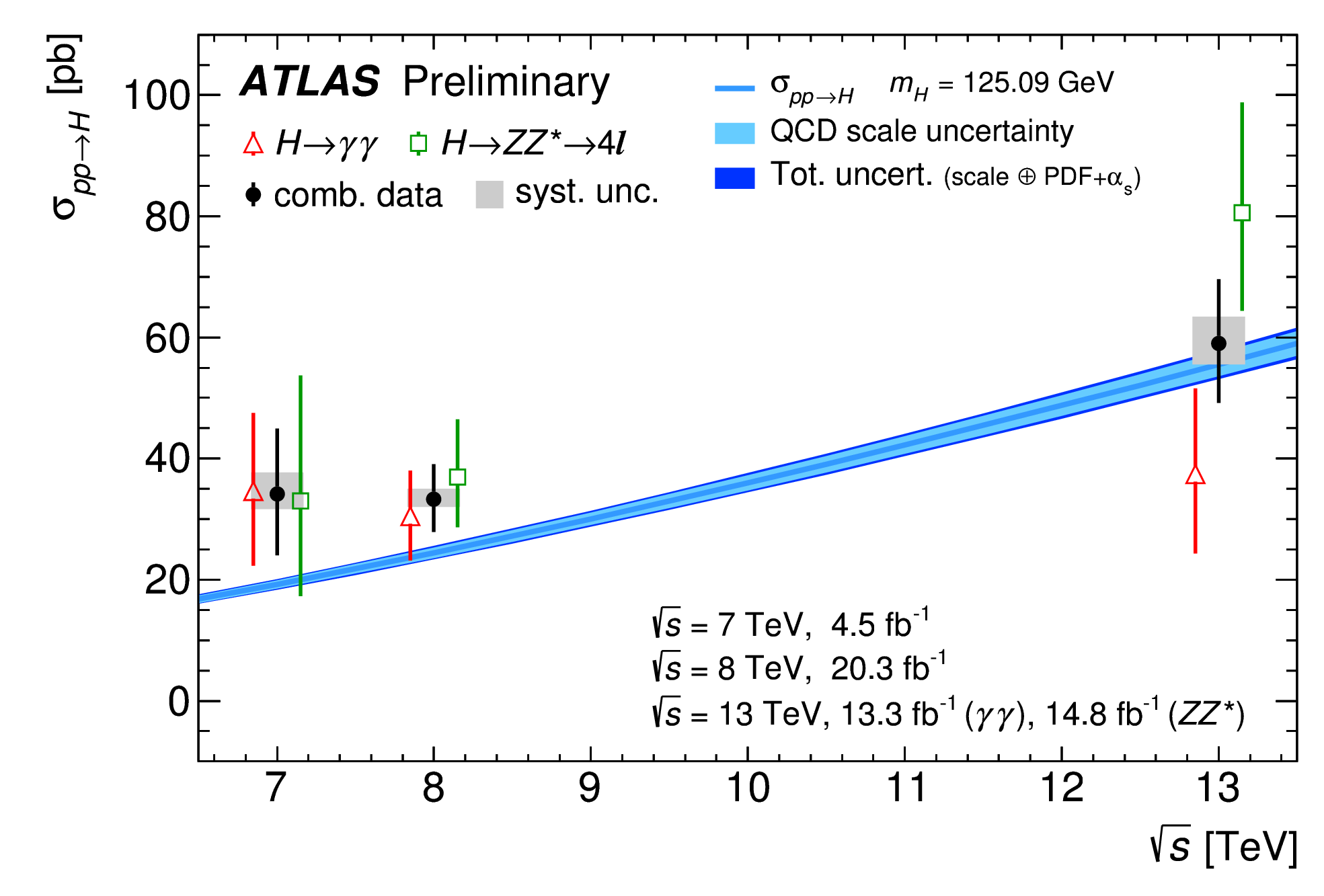}
\caption{Total $pp \rightarrow H + X $ cross-sections measured at different centre-of-mass energies compared to Standard
Model predictions at up to N$^3$LO in QCD~\cite{ATLAS-CONF-2016-081}.}
\label{fig:HsigmaRun1-2}
\end{figure}

More studies have been performed using 2015 and 2016 data at $\sqrt{s} = 13$~TeV. 
The first priority has been the ``rediscovery'' of the Higgs boson at the larger centre-of-mass energy.
The analysis~\cite{ATLAS-CONF-2016-081} is based on the measurements performed in the individual $H\rightarrow\gamma\gamma$ and $H\rightarrow ZZ^{*}$ decay channels. Higgs boson production is observed in the 13 TeV dataset with a local significance of about 10 (8.6 expected), and evidence for production via vector boson fusion is seen with a local significance of about 4 (1.9 expected).
The total $pp \rightarrow H + X $ cross-sections at centre-of-mass energies of 7, 8 and 13 TeV measured in these two decay channels  are shown in  Figure~\ref{fig:HsigmaRun1-2}, along with their combination and the comparison to theoretical predictions. 
The cross-section $\sigma$($pp \rightarrow H + X $, 13 TeV) is  $59.0^{+9.7}_{-9.2} \mathrm{(stat.)}  ^{+4.4}_{-3.5} \mathrm{(syst.)} $~pb, while the  SM prediction is $55.5^{+2.4}_{-3.4}$~pb.

The other Run 2 priorities are the refining of Higgs properties as couplings and mass; the search of  $t\bar{t}H$  production; the search for $H \rightarrow b\bar{b}$ decay (described below); the search for rare decays; the use of the Higgs boson as a tool to observe new physics.

The decay with the largest predicted branching fraction (58\%) for a SM Higgs boson of mass 
125 GeV is $H \rightarrow b\bar{b}$. However at the LHC the overwhelming backgrounds arising from multi-jet production make
a fully inclusive search extremely challenging. The production modes where the Higgs boson is produced
together with a $W$ or $Z$ boson provide a promising  alternative despite having a cross-section more than an order of magnitude lower than the dominant gluon-gluon fusion production mode. The leptonic decays of the $W$ and $Z$ boson lead to relatively clean
signatures that can be used to significantly suppress the contributions from background processes and
allow for an efficient triggering strategy.  The LHC combination of the
Run 1 ATLAS and CMS analyses resulted in observed (expected) significances of 2.6 (3.7) standard
deviations~\cite{ACcouplings}, therefore one of the main priorities for Run 2 is the measurement of the coupling with the $b$ quark. 

A search for the decay of a Standard Model Higgs boson into a $b\bar{b}$ pair when produced in
association with a $W$ or $Z$ boson has been performed using the  data 
collected in proton-proton collisions from Run 2  at a centre-of-mass
energy of 13 TeV, corresponding to an integrated luminosity of 13.2~fb$^{-1}$~\cite{ATLAS-CONF-2016-091}.
Considered final states contain 0, 1 and 2 charged leptons (electrons or muons), targeting the decays:
$Z \rightarrow \nu\nu$, $W\rightarrow l \nu$, and $Z \rightarrow ll$. For $m_H = 125$~GeV the ratio of the measured signal strength to the SM expectation is found to be  $\mu = 0.21^{+0.36}_{-0.35} \mathrm{(stat.)}\pm0.36 \mathrm{(syst.)}$. This corresponds to an observed significance of 0.42 standard deviations compared with an expected sensitivity of 1.94. The analysis procedure has been validated by measuring the yield of $(W/Z)Z$ with $Z \rightarrow b\bar{b}$,  where the ratio of the observed yield to that expected in the Standard Model was
found to be $0.91 \pm 0.17 \mathrm{(stat.)} ^{+0.32} _{-0.27} \mathrm{(syst.)}$, corresponding to a significance of 3.0 standard deviations compared to an expected significance of 3.2.

\section{Search for physics beyond the Standard Model}

A huge range of searches for physics beyond the Standard Model (BSM) have been performed
by ATLAS, looking for new sequential bosons and fermions, new vector-like quarks, signals for
extra dimensions, supersymmetric (SUSY) models, technicolour, and so on. No evidence for new 
beyond-the-Standard Model physics was observed  at either 7 or 8 TeV centre-of-mass
energy. The Run 2 dataset represents the possibility for a major extension of reach compared to Run 1 thanks to the
enhanced cross-sections at the larger energy. 
A very brief summary of some recent results is given. \\

{\bf Diboson resonances}\\
Extensions of the SM predict the existence of new particles that may
decay into vector-boson pairs, such as heavy neutral Higgs (spin 0), Heavy Vector Triplet W’ (spin 1), Bulk Randall-Sundrum Graviton G* (spin 2).  Results from the search for resonances  with masses above 1 TeV decaying to the diboson final states,
$WW$, $WZ$ and $ZZ$, in the fully-hadronic channel are reported~\cite{ATLAS-CONF-2016-055}.   Hadronic decays of the highly boosted $W$ and $Z$ bosons emerging from the decay of a heavy resonance are reconstructed within a single large-radius jet, and jet substructure properties are used to
select jets consistent with boson decays. This selection strongly suppresses the large backgrounds due
to SM multi-jet events. No significant
excess is observed in the analyzed data set and exclusion limits are set.
\\

{\bf Diphoton resonances}\\
\begin{figure}[!b]
\centering
\includegraphics[width=0.4\textwidth]{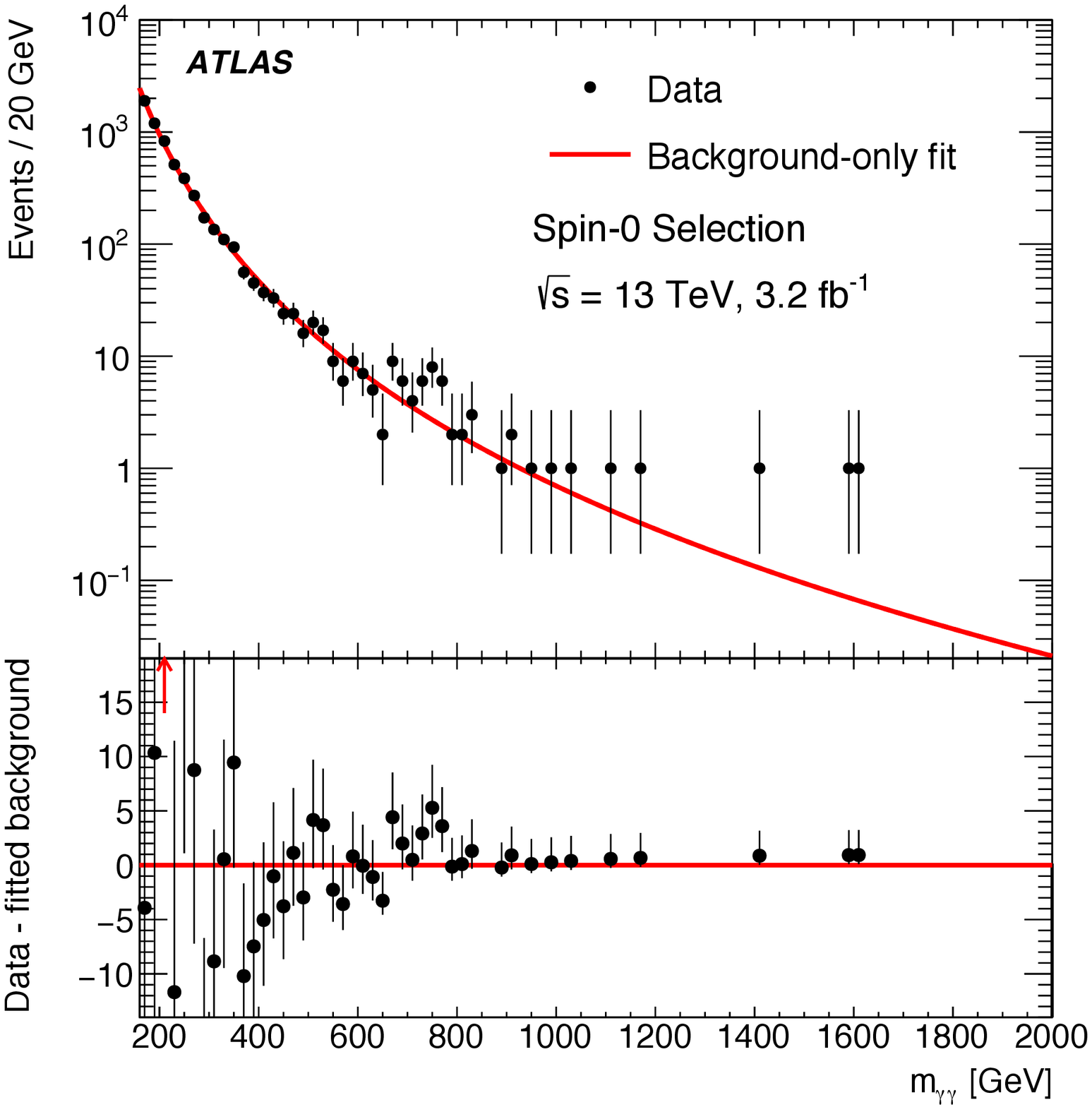}
\includegraphics[width=0.5\textwidth]{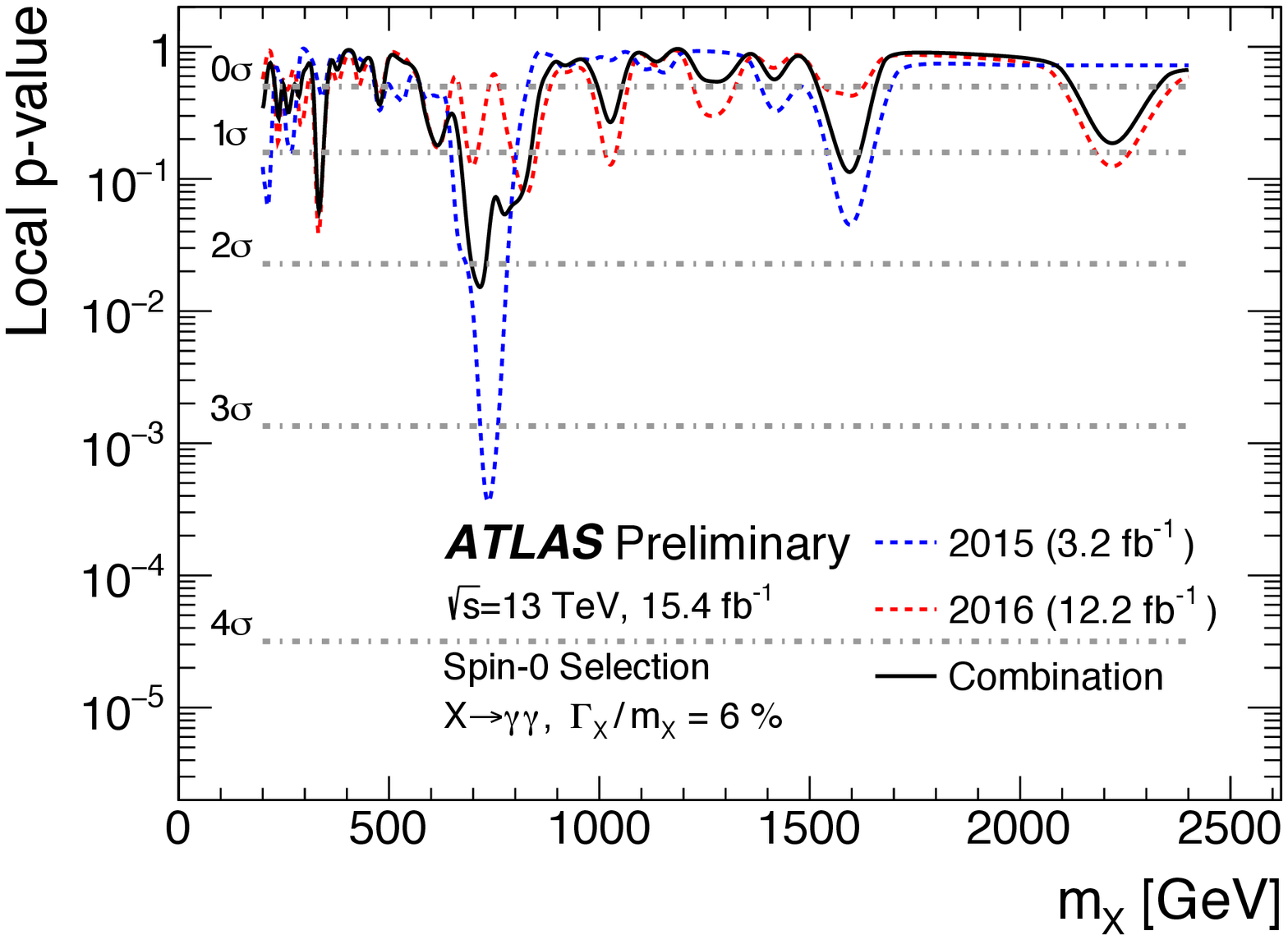}
\caption{Left: Invariant-mass distribution of the selected diphoton candidates, with the background-only fit overlaid, obtained
with 2015 data; the difference between the data and this fit is shown in the bottom panel~\cite{gammagamma2015}. Right: Compatibility with the background-only hypothesis as a function of the assumed signal mass with the full data set~\cite{ATLAS-CONF-2016-059}.
}
\label{fig:GammaGamma}
\end{figure}
Using  the 2015 data, searches for new resonances decaying into two photons observed a deviation from
the Standard Model background-only hypothesis corresponding to 3.9 standard deviations
for a resonance spin-0 mass hypothesis of 730 GeV~\cite{gammagamma2015}, see Fig.~\ref{fig:GammaGamma} left.
The excess is not confirmed in 2016 data with a four times larger statistics ~\cite{ATLAS-CONF-2016-059}:
 in the 700-800 GeV mass range the largest local significance is 2.3 standard deviations for a mass near 710 GeV and a relative width of 10\%.
The global significance of these excesses is less than one standard deviation, see Fig.~\ref{fig:GammaGamma} right.
\\

{\bf Dilepton resonances}\\
The dielectron and dimuon final-state signature has excellent sensitivity to a wide variety of new phenomena
expected in theories beyond the Standard Model. It benefits from high signal selection efficiencies
and relatively small, well-understood backgrounds. The observed dilepton
invariant mass spectrum is consistent with the Standard Model prediction, within systematic and statistical
uncertainties~\cite{ATLAS-CONF-2016-045}.  
Similarly a search for $W'$ bosons decaying to a charged lepton (electron or muon) and a
neutrino have been performed~\cite{ATLAS-CONF-2016-061}. The
transverse mass distribution is examined and no significant excess above Standard Model
predictions is observed. 
 In both cases lower limits on a resonance mass are set, enhancing the reach with respect to Run 1 of more than 1 TeV. 
\\

{\bf SUSY searches}\\
Supersymmetry (SUSY) is a generalization of space-time symmetries that predicts new bosonic partners for the
fermions and new fermionic partners for the bosons of the Standard Model and that provides a natural solution to
the hierarchy problem. The large expected cross-sections predicted for the strong production of supersymmetric particles make the production of gluinos and squarks the primary target for early searches for SUSY in $pp$
collisions at a centre-of-mass energy of 13 TeV at the LHC. \\
As an example~\cite{ATLAS-CONF-2016-095} results were reported of a search for  supersymmetric particle
production that could be observed in high-energy proton-proton collisions: events with
large numbers of jets, together with missing transverse momentum from unobserved particles,
are selected. The search selects events
with various jet multiplicities from 8 to 10 jets, and with various requirements on the
sum of masses of large-radius reclustered jets. In contrast to many other searches for the production
of strongly interacting SUSY particles in the hadronic channel, the requirement made  of large jet multiplicity implies that the threshold on missing energy can be modest. No excess above Standard Model expectations
is observed. The results are interpreted within two supersymmetry models, where gluino
masses up to 1600 GeV are excluded at 95\% confidence level, extending previous limits.

\begin{figure}[!htb]
\centering
\includegraphics[width=0.55\textwidth]{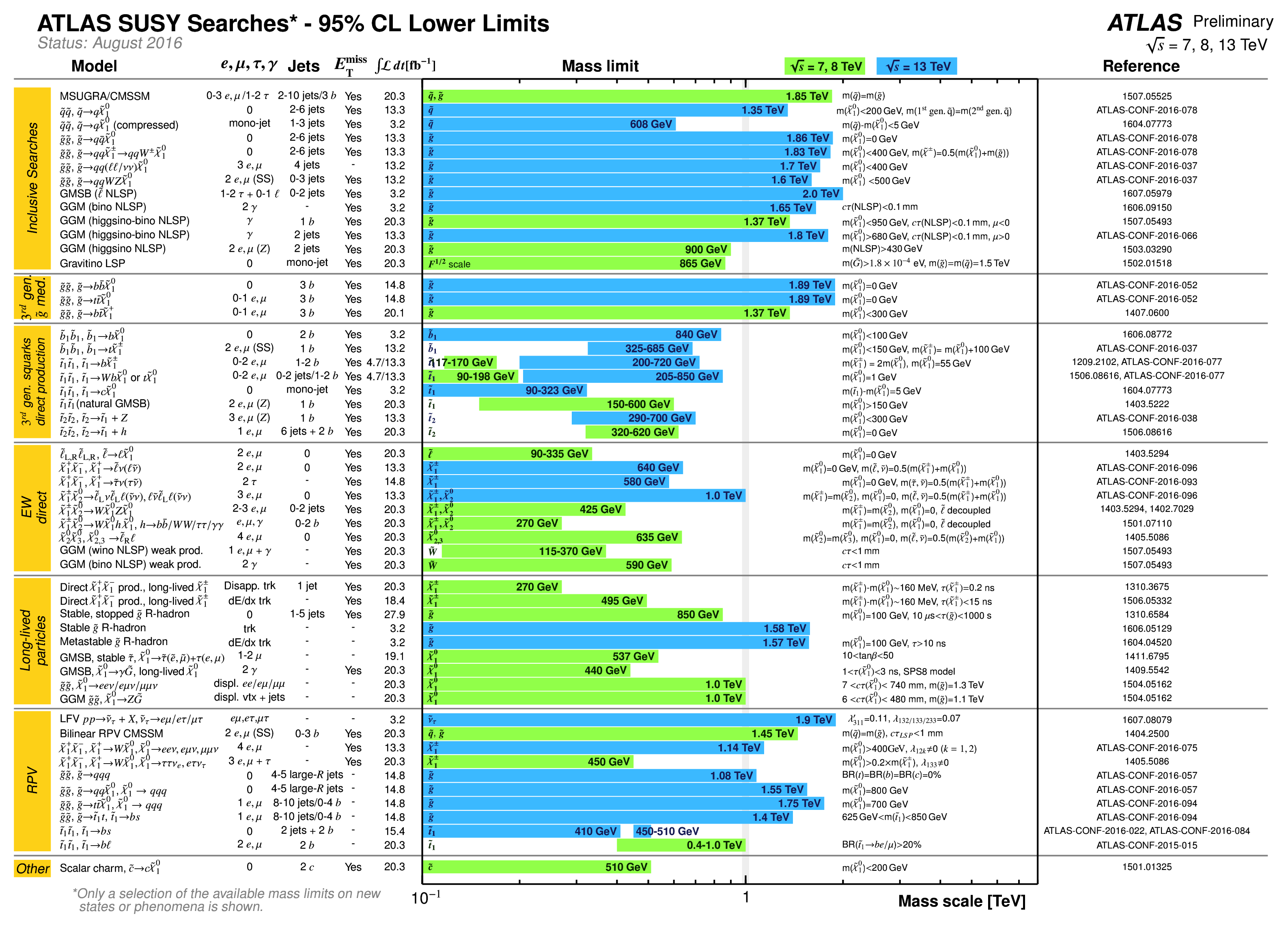}
\caption{Mass reach of ATLAS searches for Supersymmetry~\cite{SUSYPublicResults}. Only a representative selection of the available results is shown. Blue (green) bands indicate 13 TeV (8 TeV) data results.}
\label{fig:SUSY_Overview}
\end{figure}

\begin{figure}[!htb]
\centering
\includegraphics[width=0.60\textwidth]{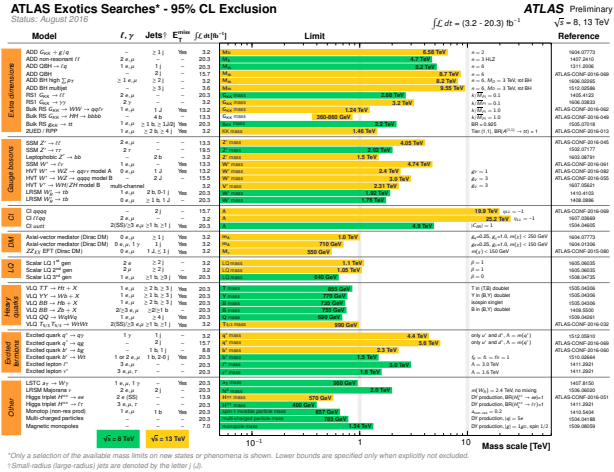}
\caption{Reach of ATLAS searches for new phenomena other than Supersymmetry~\cite{ExoticsPublicResults}. Only a representative selection of the available results is shown. Yellow (green) bands indicate 13 TeV (8 TeV) data results. }
\label{fig:Exotics_Overview}
\end{figure}

{\bf  Searches summary}\\
At the time of the conference $\sim 50\%$ of the search analyses were updated to the new Run 2 energy. In general the data agree well with the background expectations, so significant increase in excluded BSM particle mass ranges has been set. Figures~\ref{fig:SUSY_Overview} and \ref{fig:Exotics_Overview} show the reach of ATLAS searches for Supersymmetry and other new phenomena.

\section{Conclusions}
After the first LHC long shutdown ATLAS  has enhanced detectors and trigger systems that are coping very well with   a pile-up
environment beyond the design.
Many measurements of Standard Model processes have been made, accessing  simple and complex final states, probing perturbative QCD,  searching for fermionic Higgs couplings and starting precise
measurements of its properties at 13 TeV.
The larger centre-of-mass energy has allowed a major extension of reach compared to Run 1 and many topologies for BSM physics have been explored.  In general the data agree well with the background expectations, therefore a significant increase in excluded BSM particle mass ranges has been set.	
Some modest excesses are observed: the
rest of 2016 data will show if they persist or go away.

\end{document}